\documentclass[aps,pra,twocolumn,groupedaddress,amsmath,amssymb,10pt]{revtex4}
\usepackage{epsfig}
\usepackage{amssymb}
\usepackage{amsmath}
\usepackage{graphicx}
\usepackage{amsfonts}
\usepackage{epsfig}
\usepackage{revsymb}
\usepackage{bm}
\usepackage{amsmath}
\usepackage{amssymb}
\usepackage{latexsym}
\usepackage{hyperref}
\usepackage{cleveref}

\begin{document}

\title{Comment on "Zero temperature, degenerate fermion stars"
\newline
[S. Boatto, M.B. Paranjape, V. Pessanha, C.A.D. Zarro. ArXiv: 2609.20950]}
\author{Serguei N. Burmistrov} 
\affiliation{Kurchatov Institute, 123182 Moscow, Russia}
\begin{abstract} 
In contrast to S. Boatto et al. \cite{Boat}, the  degree of compactification  or compactness is clarified for the self-gravitating spheric star of fermionic matter. 
\end{abstract}
\maketitle
\par
Recently, S. Boatto et al. \cite{Boat} have reported the properties of the self-gravitating spheric star consisted of non-interacting fermions. In particular, the authors have thoroughly studied the degree of compactification $Z$ defined as 
$$
Z=\frac{R_s}{2R}=\frac{M(R)G}{c^2R}
$$
where $R_s$ is the Schwarzschild radius, $R$ is the radius of the star, $M(R)$ is the mass of the object, $G$ is the Newtonian constant of gravitation, and $c$ is the speed of light. 
\par 
The authors have described the star of radius $R$ and mass $M$ within the framework of the Newtonian gravitational potential acting on the fermion with mass $m$ as
\begin{eqnarray*}
U(r) = \left\{
\begin{array}{cc}
\frac{1}{2}G\frac{mM}{R^3}r^2 -\frac{3}{2}G\frac{mM}{R}, & r\leqslant R
\\
\\
-G\frac{mM}{r}, & r\geqslant R 
\end{array}
\right. 
.
\end{eqnarray*}
In essence the fermions are in the effective harmonic trap of radius $R$ with frequency $\omega =\sqrt{GM/R^3}$.  
 \par 
The main aspect of the study is the determination of the degree of compactification $Z$ or compactness $C(R)=2Z$  and its connection with the Buchdahl's theorem \cite{Buch}.  In this study the authors have asserted that the degree of compactification $Z$ of the self-gravitating fermion star achieves the magnitudes greater than 4/9 and exceeding the Buchdahl's bound. Thereby, there is a violation of the Buchdahl's theorem. 
\par 
The authors have given the following degree of compactification $Z$ for the number $N_s$ of fermions with the fixed spin projection    
$$
Z= \biggl(\frac{48}{71}\frac{M_p^2}{m^2}\biggr)^2\frac{1}{(36)^{1/3}(N_s)^{4/3}}\biggl[\sqrt{1+\frac{71}{12}\frac{m^4N_s^{4/3}}{M_p^4}}\, -1\biggr]^2
$$
with the limit at $N_s\gg 1$ 
$$
Z=\frac{192}{71(36)^{1/3}}.
$$
Here $M_p=\sqrt{\hbar c/G}$ is the Planck mass. 
\par 
Both the statements are erroneous. The false conclusions result from using incorrect equation (3.11) for the total binding energy $E_{\text{bind}}$ as a function of the total number $N_s$ of fermions with the fixed spin projection. Instead of the authors we must rewrite the energy of the Fermi gas in the isotropic harmonic trap with frequency $\omega$ at zero temperature, e.g. \cite{Burm},  as
 $$
\frac{1}{8}\hbar\omega (6N_s)^{4/3}. 
 $$
 \par 
Accordingly, we arrive at another result for the total binding energy $E_{\text{bind}}$ as a function of particle number $N_s$, total mass $M(R)$, and radius $R$ 
 $$
E_{\text{bind}}=\frac{\hbar\omega (6N_s)^{4/3}}{8} -\frac{3}{2}G\frac{mM}{R}N_s
 $$
where $m$ is the mass of a fermion. The total mass $M$ is determined by the equation 
 $$
 Mc^2=mc^2N_s+E_{\text{bind}} 
 $$
which solution reads
 $$
 M=\frac{32}{3}\frac{M_p^4}{m^3(6N_s)^{1/3}}\biggl[\sqrt{1+\frac{1}{32}\frac{m^4(6N_s)^{4/3}}{M_p^4}}\, -1\biggr].
  $$
The expression for the radius $R$ of the star is given by 
$$
R=\frac{9}{4}\frac{mG}{c^2}N_s\biggl[\sqrt{1+\frac{1}{32}\frac{m^4(6N_s)^{4/3}}{M_p^4}}\, -1\biggr]^{-1}.
$$
  Finally, we have for the degree of compactification  
$$
Z= \biggl(\frac{16}{3}\frac{M_p^2}{m^2}\biggr)^2\frac{1}{(6N_s)^{4/3}}\biggl[\sqrt{1+\frac{1}{32}\frac{m^4(6N_s)^{4/3}}{M_p^4}}\, -1\biggr]^2.
$$
For the limit $N_s\rightarrow\infty$,  the degree of compactification $Z$ tends to 8/9.

\end{document}